\documentstyle[aps,preprint]{revtex}
\begin{document}
\draft

\title{Magnetic and Pairing Correlations in Correlated
Electron Systems}
\author{Gang Su$^*$
}
\address{Institut f\"ur Theoretische Physik,Universit\"at zu K\"oln\\
Z\"ulpicher Strasse 77, D-50937 K\"oln, Germany
}

\maketitle
\begin{abstract}

Magnetic and superconducting pairing correlation functions 
in a general class of 
Hubbard models, the t-J model and a single-band Hubbard model with additional
bond-charge interaction are investigated, respectively. Some rigorous
upper bounds of the corresponding correlation functions are obtained. It is 
found that the decay of the spin-spin correlation function with temperature in the
general Hubbard models can not be slower than the squared inverse law at
low temperatures and the inverse law at high temperatures, while the on-site
pairing correlation function can not be slower than the inverse law. 
An upper
bound for the average energy of the t-J model is found.
The upper bounds for the spin-spin and the electron pairing correlation functions
in the t-J model as well as in the Hubbard model with 
bond-charge interaction are also obtained.
These bounds are expected to provide certain standards for approximate
methods. In some special cases, these bounds rule out the possibility of
corresponding magnetic and pairing long-range order.
    
\end{abstract}

\section{Introduction}

The study of superconductivity and magnetism in strongly correlated electron 
systems has been receiving intense interest recently. This may be attributed to
the discovery of high temperature superconductivity. Since the idea to explain
superconductivity in the framework of strongly correlated electron systems
(Hubbard model and its variants) was proposed[1], numerous investigations on
these systems have been done. For the many-body problems are difficult to 
solve, only a few rigorous results are known in literature[2]. Most of people
used approximate or numerical methods to investigate spin and pairing 
correlation functions, and usually achieved some different, 
sometimes even conflict results. In addition, some difficulties in numerical
methods remain unsolved. Therefore, a variety of interesting questions in these
systems remain ambiguous to date. 
In this situation, it is really necessary
to search for some exact results which, on the one hand, can be used to examine
the validity of some kinds of approximations or numerical calculations, and
on the other hand, can help us to further understand physical properties
of these systems. 

In this paper, we shall study magnetic and pairing correlation functions in the
following systems: a general class of Hubbard models, the t-J model and a 
single-band Hubbard model with additional bond-charge interaction. 
By using Bogoliubov's inequality, we give some
upper bounds for the studied correlation functions, 
which, at some extent,
may provide certain checks and standards for approximate methods.
 In some special cases, these
bounds rule out the possibility of the corresponding magnetic long-range order 
(LRO) and the on-site pairing LRO.

The rest of the paper is organized as follows. In Sec.II a general class of Hubbard models are studied. Symmetries of
correlation functions are discussed. The upper bounds for the magnetic correlation
function and the on-site pairing correlation function are presented, respectively.
In Sec.III and IV, the magnetic and pairing correlation functions in the t-J
model and a single-band Hubbard model with bond-charge interaction, 
are discussed, respectively. A simple upper bound for the average energy of
the t-J model is obtained. 
The upper bounds for the spin-spin correlation functions and the pairing 
correlation functions in the two 
models are given. 
In Sec.V a summary of the results are presented.

\section{Hubbard models}

Consider a general class of Hubbard models on a $d$-dimensional lattice with
$M$ (even) sites. The Hamiltonian is 
$$H = \sum_{r,r'}t_{r,r'}(a_{r}^{\dagger}a_{r'} + b_{r}^{\dagger}b_{r'})
+ \sum_{r}U_{r}a_{r}^{\dagger}a_{r}b_{r}^{\dagger}b_{r} 
- \sum_{r}\mu_{r}(a_{r}^{\dagger}a_{r} + b_{r}^{\dagger}b_{r}), \eqno(2.1)$$
where $t_{r,r'}$ is the hopping matrix element, and satisfies $t_{r,r'}^{*}
= t_{r',r}$. $a_{r}$ $(b_{r})$ annihilates a spin up (down) electron at site
$r$, $U_r $ is the local spin-independent Coulomb potential, and $\mu_r$ is
the position-dependent chemical potential. No other {\it a priori} assumption
(apart from those indicated explicitly in the context) is needed. This is thus a very general form of the Hubbard model.

Define the local spin operators as follows:
$S_{r}^{+} = a_{r}^{\dagger}b_{r}, ~~S_{r}^{-} = b_{r}^{\dagger}a_{r}, ~~
S_{r}^z = \frac{1}{2}(n_{r}^a - n_{r}^b) $
with $n_{r}^a = a_{r}^{\dagger}a_{r}$ and $n_{r}^b = b_{r}^{\dagger}b_{r}$, and
$[S_{r}^{+}, S_{r'}^{-}] = 2S_{r}^z \delta_{r,r'}, ~~[S_{r}^{\pm}, S_{r'}^z]
= \mp S_{r}^{\pm}\delta_{r,r'}. $
The global spin operators $S^{\pm} = \sum_{r}S_{r}^{\pm}$ and 
$S^z = \sum_{r}S_{r}^z$ obey the usual SU(2) symmetry and satisfy
$[S^{\pm}, H] = [S^z, H] = [S^2, H] = 0,$
where $S^2 = \frac{1}{2}(S^+S^- + S^-S^+) + (S^z)^2$ has the eigenvalue 
$S(S+1)$. The particle number $N = \sum_{r}(a_{r}^{\dagger}a_{r} + 
b_{r}^{\dagger}b_{r}) = N_{\uparrow} + N_{\downarrow}$ is conserved, and 
commutes with the Hamiltonian.
Define the $\eta$ operators as follows:
$\eta_{r}^{+} = a_{r}^{\dagger}b_{r}^{\dagger}, ~~~ 
\eta_{r}^{-} = b_{r}a_{r}, ~~~\eta_{r}^z = \frac{1}{2}(n_{r} - 1) $
with $n_{r} = n_{r}^{a} + n_{r}^{b}$, and
$[\eta_{r}^{+}, \eta_{r'}^{-}] = 2 \eta_{r}^z \delta_{r,r'}, ~~~
[\eta_{r}^{\pm}, \eta_{r'}^{z}] = \mp \eta_{r}^{\pm} \delta_{r,r'}.$
Below we will investigate the spin and the on-site pairing correlation functions
of the model, and these definitions are necessary for subsequent analyses.

\subsection{Symmetries of correlation functions}

First let us for convenience write down three well-known unitary operators 
explicitly[3], which were frequently cited in literature, and were usually 
applied to study the transformed systems connected by them, but their explicit
forms are not obvious.
 The
operator $ {\cal U}_{0} = \prod_{r} (b_r - \epsilon(r)b_{r}^{\dagger}) $
with $\epsilon(r) = (-1)^r$ and ${\cal U}_{0}^{\dagger}{\cal U}_{0}=1$ designs
the well-known particle-hole transformation[4]:
${\cal U}_{0}a_{r}{\cal U}_{0}^{\dagger}= a_{r}, ~~~
{\cal U}_{0}b_{r}{\cal U}_{0}^{\dagger}= \epsilon(r)b_{r}^{\dagger},$
which makes $U_r \rightarrow - U_r$ in the Hamiltonian (2.1) with $\mu_{r} = 
\frac{U_r}{2}$ if $t_{r,r'} =
-t$ for nearest neighbors and zero for others, like the standard single-band
Hubbard model. Here $(-1)^r$ can be understood as a factor $e^{i{\bf Q}\cdot
{\bf r}}$ with ${\bf Q}= (\pi,\pi,...)$ in two or three dimensions. The 
operator ${\cal U}_{1} = \prod_{r} (a_{r} - a_{r}^{\dagger})
(b_{r} - b_{r}^{\dagger}) $
with ${\cal U}_{1}^{\dagger}{\cal U}_{1}=1$ designs another symmetric 
particle-hole transformation:
${\cal U}_{1}a_{r}{\cal U}_{1}^{\dagger}= a_{r}^{\dagger}, ~~~
{\cal U}_{1}b_{r}{\cal U}_{1}^{\dagger}= b_{r}^{\dagger}  $
which makes $S_{r}^{+} \rightarrow - S_{r}^{-}$, $S_{r}^{z} \rightarrow - 
S_{r}^{z}$, $\eta_{r}^{+} \rightarrow - \eta_{r}^{-}$ and $t_{r,r'} \rightarrow
- t_{r',r}$ in (2.1) with additional constants and proper adjustment of the
chemical potential. The operator 
${\cal U}_{2} = exp[\frac{i\pi}{2}\sum_{r}(a_{r}^{\dagger} - b_{r}^{\dagger})
(a_{r} - b_{r})]   $
with ${\cal U}_{2}^{\dagger}{\cal U}_{2}=1$ exchanges the spins (up-down 
symmetry):
${\cal U}_{2}a_{r}^{\dagger}{\cal U}_{2}^{\dagger} = b_{r}^{\dagger},$
which leaves the Hamiltonian (2.1) unchanged, but makes $S_{r}^{+} \rightarrow
S_{r}^{-}$ and $S_{r}^{z} \rightarrow - S_{r}^{z}$. Of course, these unitary
operators can be used either individually or in a combined way.

We now give some relations for thermal correlation functions. It is noteworthy
that some of which
are trivial, but some are less obvious and thus worth to write down. As the
expectation value of the commutator, 
$[n_{r}^{a}, H] = \sum_{r'}(t_{r,r'}a_{r}^{\dagger}a_{r'} - t_{r',r}a_{r'}^{\dagger}a_{r})$, vanishes, we get
$$\sum_{r'}t_{r,r'}<a_{r}^{\dagger}a_{r'}> = \sum_{r'}t_{r',r}<a_{r'}^{\dagger}a_{r}>,  \eqno(2.2)$$
where $<\cdots>$ denotes the thermal average. Using the up-down symmetry, 
connected by ${\cal U}_{2}$, we have
$$\sum_{r'}t_{r,r'}<b_{r}^{\dagger}b_{r'}> = \sum_{r'}t_{r',r}<a_{r'}^{\dagger}a_{r}>.  \eqno(2.3)$$
Note that (2.2) and (2.3) are general, not limited to the translation
invariant case, for we have not made use of any spatial symmetry of the 
lattice. The expectation value of the commutator $[S_{r'}^{+}S_{r}^{z}, S^{-}] = 2S_{r'}^{z}S_{r}^{z}
- S_{r'}^{+}S_{r}^{-}$ yields
$$<S_{r'}^{+}S_{r}^{-}> = 2 <S_{r'}^{z}S_{r}^{z}>  \eqno(2.4)$$
due to $S^{-}$ commuting with $H$. This symmetry is essential, for it is local,
and is valid for any $r$ and $r'$. Similarly, we can obtain a lot of such 
symmetries, for instance,
$$<S_{r'}^{z}S_{r}^{\pm}> = 0, ~~~<S_{r}^{\pm}> = <S_{r}^{z}> = 0, ~~~
<S_{r}^{z}n_{r'}> = 0, $$
$$<S_{r'}^{+}S_{r}^{-}> = 2<n_{r}^{a}S_{r'}^{z}> = - 2<n_{r}^{b}S_{r'}^{z}>, 
~~~<n_{r'}^{a}S_{r}^{z}> = <n_{r}^{a}S_{r'}^{z}>,$$
$$<S_{r'}^{+}S_{r}^{-}> = 2<S_{r}^{z}S_{r'}^{+}S_{r'}^{-}> = 2<S_{r}^{+}S_{r}^{-}S_{r'}^{z}>, $$
$$<n_{r'}^{a}n_{r}^{b}> =<n_{r}^{a}n_{r'}^{b}>, ~~~<a_{r}b_{r}> = 
<b_{r}^{\dagger}a_{r}^{\dagger}> = 0,$$
$$<(S_{r}^{z})^{2m+1}> = 0   \eqno(2.5)$$
with $m=1,2,3,...$, etc. It can be seen that some of above relations are 
obvious for translation
 invariant
case, but less obvious for the system without the translation invariance,
which are probably useful for numerical calculations, and meanwhile give some 
restrictions on approximate methods. Moreover, one may apply the unitary 
transformations, as mentioned above, to Eqs.(2.5), and  can obtain the
corresponding symmetries of correlation functions for the transformed systems.
We will apply them below.

\subsection{Magnetic correlation}

Let us study the transverse spin correlation function $<S_{r'}^{+}S_{r}^{-}>$ 
for $r \neq r'$, 
which is related to the longitudinal correlation function through (2.4),
by means of Bogoliubov's inequality[5]
$$|<[A, B]>|^2 \leq \frac{\beta}{2}<\{A, A^{\dagger}\}><[[B, H], B^{\dagger}]>,
\eqno(2.6)$$
with $\beta = T^{-1}$ ($k_{B}=1$) the inverse temperature. Note that the relation between spin
correlation function and magnetic LRO has been extensively
established thirty years ago[6]. Since
$$[[S_{r}^{z}, H], S_{r}^{z}] = - \frac{1}{4} \sum_{r'(\neq r)}
(t_{r',r}a_{r'}^{\dagger}a_{r} + t_{r,r'}a_{r}^{\dagger}a_{r'} +
t_{r',r}b_{r'}^{\dagger}b_{r} + t_{r,r'}b_{r}^{\dagger}b_{r'}),$$
we have
$$<[[S_{r}^{z}, H], S_{r}^{z}]> = - \sum_{r'(\neq r)}t_{r,r'}
<a_{r}^{\dagger}a_{r'}> \geq 0, \eqno(2.7)$$
where the non-negativity of (2.7) comes from the fact that the inner 
product $(B,B^{*}) \geq 0$[5,7]. By the Schwartz inequality 
$|<A^{\dagger}B>| \leq \sqrt{<A^{\dagger}A><B^{\dagger}B>},$ 
we observe that
$$<\{S_{r'}^{+}S_{r}^{-}, S_{r}^{+}S_{r'}^{-}\}> \leq 8 ~~(r' \neq r). 
\eqno(2.8)$$
To obtain inequality (2.8) one has to substitute the definitions of spin 
operators into the left-hand side of (2.8), and then apply the Schwartz inequality to
electron operators repeatedly by noting that $|<a_{r'}^{\dagger}a_r>| \leq 1$
and $|<b_{r'}^{\dagger}b_r>| \leq 1$.
Therefore, setting $A=S_{r'}^{+}S_{r}^{-}$ ($r \neq r'$) and $B=S_r^{z}$ in 
(2.6) and noticing (2.7) and (2.8), we get a bound
$$|<S_{r'}^{+}S_{r}^{-}>|^2 \leq -4\beta \sum_{r'(\neq r)}t_{r,r'}
<a_{r}^{\dagger}a_{r'}>    \eqno(2.9)$$
for $r' \neq r$. 
One may note that the index $r'$ in the right-hand side (RHS) of (2.9) has been
eliminated due to using the Schwartz inequality in (2.8). The same situation 
occurs in the following.
Obviously, as $t_{r',r}(r' \neq r) \rightarrow 0$, then 
$|<S_{r'}^{+}S_{r}^{-}>| = 0,$ which implies that there is no spin-spin
correlation in the atomic limit, i.e., no magnetic (ferromagnetic and
antiferromagnetic) LRO occurs in this case. One may notice that the RHS of inequality (2.9) depends only on off-diagonal correlation function
$<a_{r}^{\dagger}a_{r'}>$ and hopping matrix element $t_{r,r'}$, independent
of local Coulomb potential $U_{r}$.   
 
Define operators
$$\alpha_{r,r'}^{\dagger} = t_{r',r}^{\frac12}a_{r'} - t_{r,r'}^{\frac12}a_{r}, ~~~
\alpha_{r,r'} = t_{r,r'}^{\frac12}a_{r'}^{\dagger} - t_{r',r}^{\frac12}a_{r}^{\dagger}.  \eqno(2.10)$$
Then $<\alpha_{r,r'}^{\dagger}\alpha_{r,r'}> \geq 0$ gives
$$ \sum_{r'(\neq r)}t_{r,r'}<a_{r}^{\dagger}a_{r'}> \geq \frac{1}{2}
\sum_{r'(\neq r)}|t_{r',r}|(<a_{r'}^{\dagger}a_{r'}> + <a_{r}^{\dagger}a_{r}>
-2).  \eqno(2.11)$$
Substituting (2.11) into (2.9) we have
$$ |<S_{r'}^{+}S_{r}^{-}>|^2 \leq 2\beta \sum_{r'(\neq r)}|t_{r',r}|(2 - 
<a_{r'}^{\dagger}a_{r'}> - <a_{r}^{\dagger}a_{r}>).  \eqno(2.12)$$
If the system has translation invariance, then we from (2.12) verify
 rigorously a trivial
fact $|<S_{r'}^{+}S_{r}^{-}>| = 0$ for $r' \neq r$ at full-filling. Although
the bound (2.9) is lower than (2.12), sometimes the latter is also expected
to be useful.

Since the RHS of (2.9) is intimately related to $<a_{r}^{\dagger}a_{r'}>$,
we now discuss it. For $r' \neq r$, we have
$[a_{r}^{\dagger}a_{r'}, a_{r}^{\dagger}a_{r}] = - a_{r}^{\dagger}a_{r'}.$
By setting $A = a_{r}^{\dagger}a_{r'}$ and $B = a_{r}^{\dagger}a_{r}$ in
(2.6) one obtains
$$|<a_{r}^{\dagger}a_{r'}>|^2 \leq -\beta <(n_{r}^a - n_{r'}^a)^2> 
\sum_{r'(\neq r)}t_{r,r'}<a_{r}^{\dagger}a_{r'}>
\leq - 2\beta \sum_{r'(\neq r)}t_{r,r'}<a_{r}^{\dagger}a_{r'}>$$
for $r' \neq r$. This is a recursion inequality. It turns out to be
$$|<a_{r}^{\dagger}a_{r'}>| \leq  2\beta \sum_{r'(\neq r)}|t_{r,r'}|.
\eqno(2.13)$$
Moreover, we are also able to obtain a bound
$$|<a_{r}^{\dagger}a_{r'}>| \leq \sqrt{(<n_{r}^{a}> - <n_{r}^{a}n_{r'}^{a}>)
(1-<n_{r}^{a}>)} \eqno(2.14) $$
for $ r \neq r'$. The two bounds combining (2.9) may help us to understand more about the
spin correlation function.

Let us turn to a special case for the moment. Assume that $t_{r,r'} = -t$ with $t>0$ for
$r,r'$ being nearest neighbors, and $0$ otherwise, like the standard 
single-band Hubbard model but including the local Coulomb potential $U_r$. We
introduce the Fourier transform of $a_r^{\dagger}$ as
$a_r^{\dagger} = \frac{1}{\sqrt{M}}\sum_p a_p^{\dagger}e^{-i
 p r}, $
where the summation on $p$ runs over the dual lattice defined by the boundary
conditions. By
summing over $r$ ($\neq r'$) on both sides of (2.9) and inserting the Fourier 
transform into it we obtain
$\sum_{r(\neq r')}|<S_{r'}^{+}S_{r}^{-}>|^2 \leq 2 t \beta \sum_{p,\delta}
 <n_p> e^{ip\delta},$
where $|\delta|$ denotes the lattice spacing between nearest neighbors, and
$n_p = a_{p}^{\dagger}a_{p} + b_{p}^{\dagger}b_{p}.$ One may observe that
when $<n_p> = 1$ or constant, we have $|<S_{r'}^{+}S_{r}^{-}>|=0$ for $r \neq r'$ due to
$\sum_{p}e^{ip\delta} = 0$. In other words, the single-band Hubbard model
with local Coulomb potential does not exhibit magnetic LRO at finite 
temperatures at $<n_p> = 1$ or constant. This result is independent of the sign of $U_r$
and for arbitrary dimensions. 
Although the condition of $<n_p> = 1$ or constant is very 
special, we rigorously rule out the possibility of magnetic LRO in the case.

If we donot bound $<\{S_{r'}^{+}S_{r}^{-},
S_{r}^{+}S_{r'}^{-}\}>$ by the Schwartz inequality in (2.8),  we have an
expression
$$<S_{r'}^{+}S_{r}^{-}> \leq \frac{1}{2}
<S_{r'}^{+}S_{r'}^{-}S_{r}^{+}S_{r}^{-}>, \eqno(2.15)$$
where we have used (2.5) and $<\{A^{\dagger},A\}> \geq 0$. Then from (2.6)
 we get 
$$|<S_{r'}^{+}S_{r}^{-}>| \leq \beta Q(r) + \sqrt{Q(r)^2\beta^2 + \beta P(r,r')
Q(r)}   $$
with $Q(r) = - \sum_{r'(\neq r)}t_{r,r'}<a_r^\dagger a_{r'}>$ and $ P(r,r') 
= |<S_{r'}^{+}S_{r'}^{-}S_{r}^{+}S_{r}^{-}>| \leq 1$ for $ r \neq r'$. 
By noticing
(2.13) we can obtain an upper bound for the spin-spin correlation function
$$|<S_{r'}^{+}S_{r}^{-}>| \leq G(\beta) = 2 \beta^2 R(r) + \sqrt{4 \beta^4 R(r)^2 
+ 2 \beta^2 R(r)}, \eqno(2.16)$$
with $R(r)= (\sum_{r'(\neq r)}|t_{r,r'}|)^2$. To show the bound unambiguously,
we assume that $t_{r,r'}=t$ for $r,r'$ being nearest neighbors and $0$ 
otherwise. We plot $G(\beta)$ versus temperature $\beta^{-1}$, as shown 
in Fig.1, where $\beta^{-1}$ is in units of $2t$, and the coordinate numbers are taken as $6, 4, 2$, 
respectively. Since $|<S_{r'}^{+}S_{r}^{-}>| \leq 1$, we only plot the 
interesting part. From Fig.1 one may see that the bound decreases rapidly with 
increasing temperature. When $\beta^{-1} >20$, the bound decreases slowly,
eventually to zero, as temperature increases. Evidently, the decay of the 
spin-spin
correlation function with temperature can not be slower than 
the squared inverse law at low temperatures and the inverse law at high 
temperatures. 
Although the
bound as well as the bound (2.12) can not give a general reply if the system 
can exbihit magnetic LRO or not, it
may shed useful light on examining the validity of some kinds of
approximations and numerical results, especially on the dependence of 
temperature of spin-spin correlation functions. We note that the spin-spin 
correlation function of the single-band Hubbard model was studied numerically
on small sizes ($4\times 4$) of a square lattice at half-filling at low 
temperatures[8]. 
Their calculated results 
are found to be smaller than the present bound, as indicated in Fig.1. The 
reason for this discrepancy may be that in spite of the finite-size effects 
in numerical calculation, the present bound is suitable for 
macroscopic sizes of lattices and is better for high temperatures.

\subsection{Pairing correlation}

To investigate the on-site superconducting correlation, we need to calculate
the on-site pairing correlator $<\eta_{r}^{+}\eta_{r'}^{-}> = 
<a_{r}^{\dagger}b_{r}^{\dagger}b_{r'}a_{r'}>$ in the off-diagonal long-range
(ODLR) limit[9] $|r-r'| \rightarrow \infty$, namely, off-diagonal long-range
order (ODLRO)[10]. As before, we use the Bogoliubov's inequality. Choosing $A = 
\eta_{r}^{+}\eta_{r'}^{-}$ and $B = \eta_{r}^{z}$ in (2.6), and noticing that
$[\eta_{r}^{+}\eta_{r'}^{-}, \eta_{r}^{z}] = - \eta_{r}^{+}\eta_{r'}^{-}$ for
$r \neq r'$, and
$$<[[\eta_{r}^{z}, H], \eta_{r}^{z}]> = - \sum_{r'(\neq r)}t_{r',r}
<a_{r'}^{\dagger}a_{r}>,  \eqno(2.17)$$
we have the bound
$$|<\eta_{r}^{+}\eta_{r'}^{-}>|^2 \leq - \beta \sum_{r'(\neq r)}t_{r',r}
<a_{r'}^{\dagger}a_{r}>,  \eqno(2.18)$$
where we have used the Schwartz inequality to bound $<\{\eta_{r}^{+}\eta_{r'}^{-},
\eta_{r'}^{+}\eta_{r}^{-}\}> \leq 2$. One may observe that if $t_{r,r'} 
\rightarrow 0$, then $|<\eta_{r}^{+}\eta_{r'}^{-}>| \rightarrow 0$. This 
suggests that no on-site pairing correlation in the general Hubbard model
exists in the atomic limit. 

Since 
$$ \sum_{r,r'(r\neq r')}<a_{r'}^{\dagger}a_{r}> = \sum_{r,r'}
<a_{r'}^{\dagger}a_{r}> - N_{\uparrow} = M<n_{0}^a> - N_{\uparrow}, \eqno(2.19) $$
with $<n_{0}^a> = <a_{0}^{\dagger}a_{0}>$ the number density with zero momentum
of spin-up electrons, and further assuming that $t_{r',r} \equiv t =const.$,
from (2.18) one obtains
$$ \frac{1}{M}\sum_{r}|<\eta_{r}^{+}\eta_{r'}^{-}>|^2 \leq -\frac{\beta t}{2}
(<n_{0}> - \rho),  \eqno(2.20)$$
with $<n_{0}> = <a_{0}^{\dagger}a_{0} + b_{0}^{\dagger}b_{0}>$ and $\rho =
\sum_{r}<n_r>/M$.  We note that if
$t>0$ and $<n_{0}> \geq \rho$ or $t<0$ and $<n_{0}> \leq \rho$, then 
$|<a_{r}^{\dagger}b_{r}^{\dagger}b_{r'}a_{r'}>| = 0$ for $r \neq r'$ from 
(2.20). In other words, the system can not exhibit the on-site pairing
condensation in the aforementioned circumstances. From the derivation, one
may note that the electron hopping terms plays a key role in pairing 
condensation phenomena in itinerant electron systems. Besides, one may observe
that the sign of $t$ also has the effect on the final result, as shown above.
Of course, this argument can also apply to (2.9). 

If we exchange $A$ and $B$ in the derivation of (2.18), then we can get
$|<\eta_{r}^{+}\eta_{r'}^{-}>|^2 \leq \beta <(\eta_{r}^{z})^2><[[B,H], 
B^{\dagger}]>$. One may see that if
$$<n_{r}^{a}n_{r}^{b}> \leq \frac12 (<n_{r}> - 1), \eqno(2.21)$$
it gives $|<\eta_{r}^{+}\eta_{r'}^{-}>| = 0$. Namely, under the condition of
(2.21), the Hubbard model can not appear the on-site pairing condensation at
finite temperatures. The condition (2.21) is not peculiar, e.g., 
the case of half-filling with $<n_{r}^{a}n_{r}^{b}>=0 $ falls in it.

If $t_{r,r'}= -t$ for $r,r'$ being nearest neighbors and/or next nearest 
neighbors and $0$ for others, then similar to discussions for spin correlation
functions we also have
$|<a_{r}^{\dagger}b_{r}^{\dagger}b_{r'}a_{r'}>| = 0$
for $r \neq r'$ at $<n_p> =1$. We notice that Veilleux et al[11] have recently
studied the pair correlations of the Hubbard model with next nearest neighbor
hopping by using quantum Monto Carlo method. Their consequences are 
qualitatively in
agreement with the present rigorous result in their studied parameter region.

By substituting (2.13) into (2.18), we get a rigorous upper bound for the
off-diagonal element of the two-particle reduced density matrix
$$|<a_{r}^{\dagger}b_{r}^{\dagger}b_{r'}a_{r'}>| \leq \sqrt{2} \beta 
\sum_{r'(\neq r)}|t_{r',r}|,  \eqno(2.22)$$
which should also be valid in the limit $|r-r'| \rightarrow \infty$. This bound
suggests that the decay of ODLRO with temperature is not slower than the
inverse law. Although the bound is too loose to compare with the numerical 
data on
small clusters at low temperatures[8], it gives a hint for the pairing 
correlation function in the thermodynamic limit. 
Since $<a_{r}^{\dagger}b_{r}^{\dagger}b_{r'}a_{r'}>$ is related to the
superconducting order parameter $<a_{r}b_{r}>$ (quasi-average) 
by the asymptotic 
property[12] $<a_{r}^{\dagger}b_{r}^{\dagger}b_{r'}a_{r'}> \rightarrow 
|<a_{r}b_{r}>|^2$ in the ODLR limit, (2.22) provides a standard for 
approximants in calculating the temperature-dependence of the pairing order
parameter.
Here we would like to point out that one may obtain the similar bounds as
(2.22) for other pairings, for instance, the extended s-wave, d-wave, and so 
forth[13].

\section{t-J model}

This model has been extensively studied in recent years, but the rigorous
result is rare, except that the one-dimensional (1D) supersymmetric model 
($J=\pm 2t$) is exactly solved using Bethe ansatz[14]. Many approximate or 
numerical results on magnetic and
pairing correlations in high dimensions in this model are quite different 
so far[8]. 
We study the following Hamiltonian
$$H_{t-J} = -t \sum_{<r,r'>}(a_{r}^{\dagger}a_{r'} + b_{r}^{\dagger}b_{r'})
+ J \sum_{<r,r'>}({\bf S}_{r} \cdot {\bf S}_{r'} - \frac14 n_rn_{r'}) 
\eqno(3.1)$$
on a $d$-dimensional lattice, where the notions are the same as in Sec.II,
$<r,r'>$ are nearest neighbors, $J>0$ (we here consider $J$, without loss 
of the generality, 
as an independent parameter) and $t>0$. 
The different forms of the
model has been discussed elsewhere[15]. In this model, we assume that the 
double occupancy of every site is excluded. In other words, each lattice site
is constrained to have either one electron (with spin up or down) or none, as
usual. It can be seen that the system has 
SU(2) spin symmetry. We may also obtain some symmetries of correlation
functions as in Sec.II. In this section we will first derive an upper bound for
the average energy, then
study the spin-spin
correlation function, and finally discuss the nearest neighbor pairing 
correlation function.

\subsection{Upper bound for the average energy}

From (3.1) we find
$$<[[S_{r}^{+}, H_{t-J}], S_{r}^{-}]> = t\sum_{r'_{<r>}} <a_{r'}^{\dagger}a_{r}
+ b_{r}^{\dagger}b_{r'}> - 2J \sum_{r'_{<r>}}<2S_{r}^{z}S_{r'}^{z} +
S_{r'}^{+}S_{r}^{-}>,  \eqno(3.2)$$
where $r'_{<r>}$ denotes the summation on $r'$ running over nearest neighbors
of $r$. (3.2) then implies
$$\sum_{r'_{<r>}}<S_{r'}^{+}S_{r}^{-}> \leq \frac{t}{4J}\sum_{r'_{<r>}}
<a_{r'}^{\dagger}a_{r} + b_{r}^{\dagger}b_{r'}>, \eqno(3.3)$$
where we have used (2.4) and the non-negative property[5,7] of (3.2).
By noticing (3.3) and $ <S_{r}^{+}S_{r'}^{-}> = <S_{r'}^{+}S_{r}^{-}>$ 
one gets
$<H_{t-J}> \leq - \frac{5}{8}t\sum_{\delta, p}<n_p> e^{ip\delta}$ $ - \frac{J}{4}
\sum_{<r,r'>}<n_{r}n_{r'}>. $
At temperature $T$, we have the average energy (internal energy) $E_{0} = <H_{t-J}>$. 
On the other hand, the non-negativity of 
$<[[a_rb_r, H_{t-J}], b_r^{\dagger}a_r^{\dagger}]>$ yields $ -t\sum_{<r,r'>}
<a_{r}^{\dagger}a_{r'} + b_{r}^{\dagger}b_{r'}> \leq \frac{J}{2}\sum_{<r,r'>}
<n_rn_{r'}> - \frac{J}{2}Nz$ with $z$ the coordinate number, and thus
$E_{0} \leq  \frac{J}{16}\sum_{<r,r'>}<n_rn_{r'}> - \frac{5J}{16}Nz$. 
Furthermore, since $<(n_r - n_{r'})^2> \geq 0$ and noting that $<n_r^an_r^b>=0$
due to the restriction of no doubly occupied sites, one has $<n_rn_{r'}>
\leq \frac12 (<n_r> +<n_{r'}>)$. Substituting these facts into $E_{0}$ we have
$$E_{0} \leq -\frac{1}{4}JzN. \eqno(3.4)$$
We like to point out that the bound (3.4) is generic, not limited to the 
translation invariant system, and is valid for arbitrary filling fraction and
arbitrary dimensions. 
If the system has the singlet ground state, then $E_{0}$ at $T \rightarrow 0$
can be regarded as the ground state
energy, and $<...>$ thus means the average in the ground state. 
If the 
ground state of the system is degenerate, $E_{0}$ can also be understood as
the ground-state energy for all ground states. At half-filling, the t-J model 
reduces to the Heisenberg antiferromagnetic model. In 1D the ground state
energy is well known to be $E_0/M = -Jln2$[14], which clearly satisfies (3.4):
$E_0/M = -0.693147J < -0.5J$.
 Away from the half-filling, the ground state energy of the supersymmetric t-J 
model also comply the bound (3.4), as shown in Bares et al's paper[14]. In 2D,
the numerical result for estimating the ground 
state energy of one hole in the interval $0.2\leq J/t \leq 1.0$ 
on small clusters ($4 \times 4$) is $E_0/M =
-3.17t + 2.83t(J/t)^{0.73}$ [16]. However, their Hamiltonian does not contain
the term $-\frac{J}{4} \sum_{<r,r'>}n_rn_{r'}$. If this term is taken into
account, their numerical results would be in agreement with the present bound. Since this bound
is a rigorous result, it can be expected to provide checks for 
approximate and numerical methods, especially in high dimensions.

\subsection{Magnetic correlation}
 
The magnetic structure factor $m(p)$ is given by
$$m(p) =\frac{1}{M}\sum_{r,r'}e^{ip(r-r')}<S_{r}^{z}S_{r'}^{z}> = <S_{p}^{z}
S_{-p}^{z}>,   \eqno(3.5)$$
where $S_{p}^z = \frac{1}{\sqrt{M}}\sum_{r}S_r^z 
e^{ipr}$.
By symmetry, 
$<S_{p}^{z}S_{-p}^{z}> = \frac{1}{2}<S_{p}^{+}S_{p}^{-}>$. From
(3.3) we obtain
$$\sum_{p}[m(p) - \frac{t}{8J}<n_{p}>]\gamma_{p} \leq 0  \eqno(3.6)$$
with $\gamma_{p}=\sum_{\delta}e^{ip\delta}$. Note that $m(p) \geq 0$. 
Inequality (3.6) gives a severe restriction on $m(p)$ in the t-J model. If
$<n_p>=1$ or $t=0$, we have
$ \sum_{p} m(p) \gamma_{p} \leq 0. $
Since the existence of N\'eel order corresponds to $m(p)$ containing a $\delta$
function at ${\bf Q}=(\pi, \pi, ...)$ in the infinite-volume limit[17], and let $m^2$
be the coefficient of this delta function, we from (3.6) get a bound
$$m^2 \gamma_{Q} \leq \frac{t}{8J}\sum_{p}<n_p>\gamma_{p} -\sum_{p \neq Q} 
m(p) \gamma_{p}.  \eqno(3.7)$$
It has been shown that the 3D half-filled t-J model, i.e., Heisenberg 
antiferromagnetic model, has LRO[17]. Away from half-filling, (3.7) may shed 
some light on the antiferromagnetic order of the t-J model.
For a square lattice, $\gamma_{Q} = -4$. Then we have
$$m^2 \geq \frac14 \sum_{p \neq Q} m(p) \gamma_{p} - \frac{t}{32J}\sum_{p}<n_p>\gamma_{p}. $$
If we obtain a bound for $\sum_{p \neq Q} m(p) \gamma_{p}$, then we can say 
something about the antiferromagnetic LRO in the model, which will be left
for future study.

We choose $A = S_{r'}^{-}S_{r}^{z}$ and $B=S_{r}^{+}$ in (2.6). Then 
$[A,B]=S_{r'}^{-}S_{r}^{z}$ for $r \neq r'$. Since $\frac{1}{2}<\{S_{r'}^{-}
S_{r}^{z}, S_{r}^{z}S_{r'}^{+}\}> \leq \frac14$ by the Schwartz inequality, 
from (2.6) and (3.2) we
have
$$|<S_{r'}^{+}S_{r}^{-}>| \leq \sqrt{\frac{\beta}{8}(t\sum_{r'_{<r>}} 
<a_{r'}^{\dagger}a_{r} + b_{r}^{\dagger}b_{r'}> - 4J 
\sum_{r'_{<r>}}<S_{r'}^{+}S_{r}^{-}>)}$$
$$ \leq \sqrt{\frac{z\beta}{4}(t + 2J\sqrt{\frac{z\beta}{4}(t + 2J\sqrt{\frac{z\beta}{4}(t + 2J\sqrt{...})})})}.  \eqno(3.8)$$
This inequality gives an upper bound for the spin-spin correlation function 
in the t-J model. 
Particularly, as $t=0$, the model reduces to the Heisenberg
antiferromagnetic model, and (3.8) becomes
$$|<S_{r'}^{+}S_{r}^{-}>| \leq \frac{J\beta}{2}z. \eqno(3.9)$$
That is, the temperature-dependence of the spin-spin correlation function can 
not be slower than the inverse law in the Heisenberg antiferromagnetic model. 

\subsection{Pairing correlation}

Now we come to discuss the pairing correlation function. Since there is no 
doubly occupied sites in the system, the on-site pairing correlation should be vanishing. We in the following cosider the nearest neighbor pairing order
parameter $<a_{r_1}b_{r_2}>$ with $r_1,r_2$ being nearest neighbor sites. For
this purpose, we have to add a U(1) symmetry breaking term $H'= -\alpha
\sum_{<r_1,r_2>}(a_{r_1}b_{r_2} + b_{r_2}^{\dagger}a_{r_1}^{\dagger})$ as well
as the chemical potential term $-\mu \sum_{r}n_r$ into the Hamiltonian 
$H_{t-J}$. Let $B=a_{r_1}b_{r_2}$, and $A = a_{r_1}^{\dagger}a_{r_1} + 
b_{r_2}^{\dagger}b_{r_2}$. Then $[A,B] = -2a_{r_1}b_{r_2}$, $<[[a_{r_1}b_{r_2},
H_{t-J}], b_{r_2}^{\dagger}a_{r_1}^{\dagger}]>  \leq D(r_1,r_2)$ with 
$D(r_1,r_2) = t (\sum_{r_{<r_1>}} <a_{r_1}^{\dagger}a_{r}> + \sum_{r_{<r_2>}} <b_{r_2}^{\dagger}b_{r}>) - (J+2\mu) (1 - <n_{r_1}^a> - <n_{r_2}^b>) + 2J <n_{r_1}^an_{r_2}^b> + 6Jz - \alpha (\sum_{r_{<r_2>}}a_{r}^{\dagger}b_{r_2}^{\dagger}
+ \sum_{r_{<r_1>}}b_{r}^{\dagger}a_{r_1}^{\dagger})$, and $ <\{A,A^{\dagger}\}> \leq 8$. Substituting these results into (2.6) we obtain the bound
$$|<a_{r_1}b_{r_2}>|^2 \leq \beta D(r_1,r_2)  \eqno(3.10)$$
as $\alpha \rightarrow 0$ in the thermodynamic limit, where we have used the
Schwartz inequality to bound those terms with four creation and annihilation
operators. One may note that $<a_{r_1}b_{r_2}>$ in (3.10) is the Bogoliubov's quasi-average[12]. On the other hand, according to Bogoliubov's argument[12], the
off-diagonal element of the two-particle reduced density matrix $<b_{r_2'}^{\dagger}a_{r_1'}^{\dagger}a_{r_1}b_{r_2}>$ has the asymptotic behavior
$$<b_{r_2'}^{\dagger}a_{r_1'}^{\dagger}a_{r_1}b_{r_2}> \rightarrow <b_{r_2'}^{\dagger}a_{r_1'}^{\dagger}><a_{r_1}b_{r_2}>  \eqno(3.11)$$
in the ODLR limit[13] $|(r_2',r_1') - (r_2,r_1)| \rightarrow  \infty$. It is
worth mentioning that (3.11) is not incompatible with Haag's spatial cluster theorem[19].
In the
translation invariant system, the temperature-dependence of ODLRO should thus
obey
$$<b_{0}^{\dagger}a_{1}^{\dagger}a_{r_1}b_{r_2}> \leq \beta D(r_1,r_2),  \eqno(3.12)$$
i.e., the decay of the ODLRO for nearest-neighbor pairs with temperature in
the t-J model with translation invariance can not be slower than the inverse
law. This bound thus offers a check for some approximate results on the 
temperature-dependence of superconducting order parameter.

\section{Hubbard model with bond-charge interaction}

The Hamiltonian of the model is given by
$$ H_{b-c} = -t \sum_{<r,r'>}(a_{r}^{\dagger}a_{r'} + b_{r}^{\dagger}b_{r'} + 
h.c.) + U \sum_{r}a_{r}^{\dagger}a_{r}b_{r}^{\dagger}b_{r}$$
$$ + X \sum_{<r,r'>}[(a_{r}^{\dagger}a_{r'} + h.c.)(n_{r}^{b} + n_{r'}^{b})
+(b_{r}^{\dagger}b_{r'} + h.c.)(n_{r}^{a} + n_{r'}^{a})] - \mu \sum_{r}
(a_{r}^{\dagger}a_{r} + b_{r}^{\dagger}b_{r}),  \eqno(4.1)$$
where $X$ is the bond-charge interaction, and other notations are standard, as
usual.  This model has been extensively discussed by Hirsch[20], Bariev et al[21], de Boer et al
[22] and Schadschneider[23]. It has been shown[22] that the $\eta$ pairing 
state with ODLRO is the ground state if $U \leq -2Z|t|$, and can be solved 
exactly in one dimension[21-23], as $t=X$ and $\mu = U/2$. For certain values of $X$ and large
densities of electrons (small doping) the bond-charge interaction may lead to
an attractive effective interaction between the holes within the framework
of BCS mean-field theory[20]. In this section, we will give a few rigorous
bounds for the spin-spin and the on-site pairing correlation functions of this system. 

Like the way exploited in preceding sections, it is not difficult to obtain
the following two bounds
$$|<S_{r}^{+}S_{r'}^{-}>| \leq [8\beta(t \sum_{r'_{<r>}}<a_{r'}^{\dagger}a_{r}>
- X\sum_{r'_{<r>}}<a_{r'}^{\dagger}a_{r}(n_{r}^{b} + n_{r'}^{b})>)]^{\frac12}
\eqno(4.2)$$
and
$$|<a_{r}^{\dagger}b_{r}^{\dagger}b_{r'}a_{r'}>| \leq [2\beta (t \sum_{r'_{<r>}}<a_{r'}^{\dagger}a_{r}>
- X\sum_{r'_{<r>}}<a_{r'}^{\dagger}a_{r}(n_{r}^{b} + n_{r'}^{b})>)]^{\frac12}
\eqno(4.3)$$
for $ r \neq r'$ by setting $A=S_{r}^{+}S_{r'}^{-}$ and $\eta_{r}^{+}
\eta_{r'}^{-}$ with $r \neq r'$ and $B=S_{r}^{z}$ and $\eta_{r}^{z}$ in (2.6),
respectively, where we have used the spin-flip symmetry (${\cal U}_{2}$). Evidently, if $X$ satisfies the 
following condition
$$ X\sum_{r'_{<r>}}<a_{r'}^{\dagger}a_{r}(n_{r}^{b} + n_{r'}^{b})> \geq
t \sum_{r'_{<r>}}<a_{r'}^{\dagger}a_{r}>,   \eqno(4.4)$$
then there is no spin-spin correlation and on-site pairing correlation.  To
assure the existence of magnetism and superconductivity in the system 
$H_{b-c}$, the condition in opposite direction of inequality (4.4)
must be hold, which gives a restriction on the values of $X$. 

For a special case $t=X$, the system possesses the symmetric particle-hole
symmetry, connected by the unitary operator ${\cal U}_{1}$, as discussed in 
Ref.[23], the $\eta$ pairing symmetry[22] at half-filling, and so forth. This
model is really interesting, and the details will be presented in a 
seperate publication.

\section{Summary}

We have rigorously investigated magnetic and superconducting pairing 
correlation
functions in a general class of Hubbard models, the t-J model and a single-band
Hubbard model with additional bond-charge interaction by means of the
Bogoliubov's inequality, respectively. Some corresponding upper bounds are
obtained, which are expected to provide certain checks and standards for
approximate methods. In some special cases, these bounds rule out the
possibility of corresponding magnetic and pairing LRO.

For the Hubbard models, we obtained an upper bound for the spin-spin correlation
function, which indicates that the decay of the correlation function with
temperature can not be slower than the squared inverse law at low temperatures
and the inverse law at high temperatures.
From these bounds we observe that there is no 
magnetic LRO in the
atomic limit and in the case with the momentum distribution function being
constants for the model with only nearest neighbor hoppings. An upper bound
for the on-site pairing correlation function was obtained, which suggests
that the decay of the on-site pairing ODLRO with temperature is not slower than the inverse law.
Since our method is rigorous, the present result may be applied to clarify
some contradication in approximate calculations.
In addition, we found that there is no on-site pairing correlation in the 
atomic limit and in the cases either $t>0$ and $<n_{0}> \geq \rho$ or $t<0$ 
and $<n_{0}> \leq \rho$ (see (2.20)) or $<n_{p}>$ being constants for the 
single-band model but with local Coulomb interaction. 
It is emphasized that
all obtained bounds are independent of the local on-site Coulomb interaction
and are valid for arbitrary dimensions.

For the t-J model, we obtained an upper bound for the 
average 
energy (internal energy) for arbitrary electron fillings. Whenever the ground 
state of the system is unique or not, the upper bound at $T \rightarrow 0$ can be regarded as
 that of the ground-state energy. Since the bound is rigorous, it provides a standard for 
approximate and numerical methods. We also obtained a lower bound for 
the N\'eel order, which may 
shed useful light on the antiferromagnetic order of the system.
 An upper bound for the 
spin-spin correlation function was derived, which implies that the decay of it
with temperature in the 
model is not slower than the $\beta^{\frac12}$ law away from half-filling and
the inverse law at half-filling. 
An upper bound for the 
nearest-neighbor pairing correlation was obtained for the translation invariant
system, which suggests that the decay of ODLRO with temperature can not be 
slower than the inverse law.
The results hold for arbitrary dimensions.

For the Hubbard model with bond-charge interaction, we obtained two bounds for
the spin-spin correlation function and the on-site pairing correlation function, which
gives a severe restriction on the values of the bond-charge interaction.

\acknowledgments

The author has benefited from discussions with Dr. A. Schadschneider and
Prof. B.H. Zhao. He is also grateful to Prof. J. Zittartz, Dr. A. Kl\"umper and ITP of Universit\"at zu 
K\"oln for warm hospitality. This work was supported by the Alexander von Humboldt 
Stiftung.

\references
\begin{description}

\item $^*$ On leave from Graduate School, 
Chinese Academy of Sciences, Beijing, China.

\item[[1]] P.W. Anderson 1987 Science 235, 1196; F.C. Zhang and T.M. Rice 1988
Phys. Rev. B37, 3759.

\item[[2]] See {\it The Hubbard Model: Recent Results}, ed. M. Rasetti ( World
Scientific, Singapore, 1991); E.H. Lieb, in Proceedings of the conference 
``{\it  Advances in Dynamical Systems and Quantum Physics}'', 
eds. V. Figari et al(World
Scientific, Singapore, 1995); R. Strack and D. Vollhardt 1995 J. Low. 
Tem. Phys. 99, 385. 

\item[[3]] C.N. Yang and S.C. Zhang 1990 Mod. Phys. Lett. B4, 759.

\item[[4]] F.C. Pu and S.Q. Shen 1994 Phys. Rev. B50, 16086.

\item[[5]] N.D. Mermin and H. Wagner 1966 Phys. Rev. Lett. 17, 1133;
D.K. Ghosh 1971 Phys. Rev. Lett. 27, 1584.

\item[[6]] R. Griffiths 1966 Phys. Rev. 152, 240.

\item[[7]] F.J. Dyson, E.H. Lieb and B. Simon 1978 J. Stat. Phys. 18, 335.

\item[[8]] E. Dagotto 1994 Rev. Mod. Phys. 66, 763 and references therein.

\item[[9]] H.T. Nieh, G. Su and B.H. Zhao 1995 Phys. Rev. B51, 3076.

\item[[10]] C. N. Yang 1962 Rev. Mod. Phys. 34, 694.

\item[[11]] A.F. Veilleux et al 1995 Phys. Rev. B52, 16255.

\item[[12]] N.N. Bogoliubov 1960 Physica 26, S1.

\item[[13]] D.J. Van Harlingen 1995 Rev. Mod. Phys. 67, 515.

\item[[14]] P. Schlottman 1987 Phys. Rev. B36, 5177; P.A. Bares and G. 
Blatter 1990 Phys. Rev. Lett. 64, 2567.

\item[[15]] G. Su 1993 J. Phys. A26, L139.

\item[[16]] E. Dagotto, A. Moreo, R. Joynt, S. Bacci and E, Gagliano 1990 Phys. 
Rev. B41, 2585.

\item[[17]] T. Kennedy, E.H. Lieb and B.S. Shastry 1988 J. Stat. Phys. 53, 
1019.

\item[[18]] N. Kawakami and S.-K. Yang 1991 J. Phys. CM 3, 5983.

\item[[19]] R. Haag, {\it Local Quantum Physics} (Springer, Berlin, 1992).

\item[[20]] J.E. Hirsch 1989 Phys. Lett. A134, 451; 1991 Phys. Rev. B43, 
11400.

\item[[21]] R.Z. Bariev, A. Kl\"umper, A. Schadschneider, and J. Zittartz 1993
J. Phys. A26, 1249; 4863.

\item[[22]] J. de Boer, V.E. Korepin and A. Schadschneider 1995 Phys. Rev. Lett.
74, 789; J. de Boer and A. Schadschneider 1995 Phys. Rev. Lett. 75, 4298.

\item[[23]] A. Schadschneider 1995 Phys. Rev. B51, 10386, and references 
therein.

\end{description}

\noindent{\bf Figure Caption}

 Fig.1  The temperature dependence of the bound (2.16), where $\beta^{-1}$ is 
in units of $2t$, and coordinate numbers are taken as $6, 4, 2$ respectively, as indicated. The numerical data are taken from Ref.[8]. 

\end{document}